\documentclass[letter]{aa} 

\usepackage{graphicx}
\usepackage{txfonts}
\usepackage{natbib}
\titlerunning{Test of a revised prescription for angular momentum transport by the Tayler instability}

\begin{document}

   \title{Asteroseismology of evolved stars to constrain the internal transport of angular momentum}

   \subtitle{II. Test of a revised prescription for transport by the Tayler instability}

\author{P. Eggenberger\inst{1}
\and J.W. den Hartogh\inst{2}
\and G.~Buldgen\inst{1}
\and G.~Meynet\inst{1}
\and S.J.A.J.~Salmon\inst{3}
\and S. Deheuvels\inst{4}}

\institute{Observatoire de Gen\`eve, Universit\'e de Gen\`eve, 51 Ch. des Maillettes, CH-1290 Sauverny, Suisse \\ 
\email{patrick.eggenberger@unige.ch}
\and
Konkoly Observatory, Research Centre for Astronomy and Earth Sciences, Konkoly Thege Mikl\'{o}s \'{u}t 15-17, H-1121 Budapest, Hungary
\and
STAR Institute, Universit\'e de Li\`ege, All\'ee du Six Ao\^ut 19C, B-4000 Li\`ege, Belgium
\and
IRAP, Universit\'e de Toulouse, CNRS, CNES, UPS, Toulouse, France
}
   \date{Received; accepted}

 
  \abstract
   {Asteroseismic observations enable the characterisation of the internal rotation of evolved stars. These measurements reveal that an unknown efficient angular momentum (AM) transport mechanism is needed for subgiant and red giant stars in addition to hydrodynamic transport processes. A revised prescription for AM transport by the magnetic Tayler instability has been recently proposed as a possible candidate for such a missing mechanism.}
   {We compare the rotational properties predicted by this magnetic AM transport to asteroseismic constraints obtained for evolved stars with a particular focus on the subgiant phase.
}
   {We computed models accounting for the recent prescription for AM transport by the Tayler instability with the Geneva stellar evolution code for subgiant and red giant stars, for which an asteroseismic determination of both core and surface rotation rates is available.}
   {The revised prescription for the transport by the Tayler instability leads to low core rotation rates after the main sequence that are in better global agreement with asteroseismic measurements than those predicted by models with purely hydrodynamic processes or with the original Tayler-Spruit dynamo. A detailed comparison with asteroseismic data shows that the rotational properties of at most two of the six subgiants can be correctly reproduced by models accounting for this revised magnetic transport process. This result is obtained independently of the value adopted for the calibration parameter in this prescription. We also find that this transport by the Tayler instability faces difficulties in simultaneously reproducing asteroseismic measurements available for subgiant and red giant stars. The low values of the calibration parameter needed to correctly reproduce the rotational properties of two of the six subgiants lead  to core rotation rates during the red giant phase that are too high. Inversely, the higher values of this parameter needed to reproduce the core rotation rates of red giants lead to a very low degree of radial differential rotation before the red giant phase, which is in contradiction with the internal rotation of subgiant stars.
}
{In its present form, the revised prescription for the transport by the Tayler instability does not provide a complete solution to the missing AM transport revealed by asteroseismology of evolved stars.}


   \maketitle
%

\section{Introduction}
\label{intro}

Direct observational constraints on the internal rotation of stars are needed to progress in the modelling of angular momentum (AM) transport in stellar interiors. These valuable constraints are now available thanks to the capability of asteroseismic techniques to reveal internal stellar properties. Mixed oscillation modes, which are simultaneously sensitive to the properties in the core and  the external layers of a star, are of particular interest in this context. Measurements of rotational splittings of mixed modes for post-main sequence (poMS) stars have then been used to characterise the internal rotation of evolved stars \cite[][]{bec12, deh12, mos12, deh14, deh15, dim16, deh17, geh18, dim18}. 

Predictions of rotating stellar models can then be compared to these observational constraints. An initial key result based on these comparisons is that models of red giants computed by accounting only for hydrodynamic transport processes exhibit high core rotation rates, which strongly contrad with the modest degree of radial differential rotation deduced from rotational splittings of mixed modes \citep{egg12_rg,cei13,mar13}. This shows that AM transport by the shear instability and meridional currents is insufficient to correctly reproduce asteroseismic data, and that an additional efficient AM transport mechanism is at work in the radiative zones of evolved stars.

Another result of these comparisons is that the efficiency of AM transport\footnote{Efficiency of AM transport refers in this work to the effective viscosity corresponding to the additional transport process.} during the poMS can be precisely determined from asteroseismic measurements. Importantly, such a characterisation can be performed independently from all uncertainties regarding the modelling of rotational effects before the poMS evolution (e.g. for the efficiency of AM transport during the main sequence (MS) or surface magnetic braking) for both red giant \citep{egg12_rg,egg17} and subgiant stars \citep[][Paper I hereafter]{egg19}. Key trends can then be determined for internal AM transport during the poMS. First, the efficiency of the missing poMS AM transport mechanism is found to increase with the stellar mass during both the red giant \citep{egg17} and the subgiant phase (Paper I). Second, the efficiency of the additional transport mechanism is found to increase when the star ascends the red giant branch \citep[][]{can14,spa16}, while this transport efficiency decreases with the evolution during the subgiant phase (Paper I). These trends have to be reproduced by any AM transport candidate aiming at explaining the missing poMS transport efficiency.

Magnetic AM transport processes are prime candidates for ensuring an efficient coupling in stellar interiors. A first possibility is to invoke magnetic torques to ensure uniform rotation in radiative zones together with radial differential rotation in convective envelopes as proposed by \cite{kis15}. Preliminary results however indicate that rigid rotation in the radiative interior of evolved stars is disfavoured by asteroseismic measurements \citep{deh14,dim16,kli17}. Alternatively, a magnetic transport mechanism based on the Tayler instability \citep{tay73} and the winding up of an initial weak field by differential rotation has been proposed by \cite{spr02}. Interestingly, this process (known as the Tayler-Spruit dynamo) predicts an efficient AM transport in radiative zones that leads to a solar rotation profile in good agreement with helioseismic constraints \citep{egg05_mag,egg19_sun}. During the poMS evolution, this mechanism predicts however an AM transport efficiency that is insufficient to account for the low core rotation rates of evolved stars as determined from asteroseismic measurements \citep{can14,den19}. 

Recently, \citet{ful19} proposed a revised prescription for AM transport by the Tayler instability. Interestingly, this revised prescription predicts a more efficient AM transport than the original Tayler-Spruit dynamo, which could result in core rotation rates of evolved stars in better agreement with asteroseismic constraints \citep{ful19}. We note that the differences in AM transport efficiency between the prescriptions proposed by \citet{ful19} and \cite{spr02} are already visible on the  MS; there is a significant impact on the core rotation rate predicted for the Sun \citep{egg19_sun}. Owing to its capability of efficiently transporting AM, the expression for the transport by the Tayler instability proposed by \citet{ful19} constitutes a promising candidate for the missing poMS transport process. The key question is to determine whether this mechanism is able to reproduce the asteroseismic constraints mentioned above. In particular, we seek to find out whether the magnetic transport process recently proposed by \citet{ful19} is compatible with the internal rotation of subgiant stars as deduced from asteroseismic measurements. 

The input physics used to compute subgiant and red giant models that take into account both hydrodynamic and magnetic AM transport processes are described in Sect.~\ref{inputphys}. The rotational properties of these models are compared to asteroseismic measurements in Sect.~\ref{results}. The conclusion is given in Sect.~\ref{conclusion}.

\section{Input physics of the models}
\label{inputphys}

Models of subgiant and red giant stars are computed with the Geneva stellar evolution code \citep{egg08} using the assumption of shellular rotation \citep{zah92}. The internal AM transport is then followed simultaneously to the evolution of the star by taking into account meridional circulation, shear instability, and AM transport by the magnetic Tayler instability as proposed by \cite{ful19}. The following equation is then solved for AM transport in radiative zones:
\begin{equation}
  \rho \frac{{\rm d}}{{\rm d}t} \left( r^{2}\Omega \right)_{M_r} 
  =  \frac{1}{5r^{2}}\frac{\partial }{\partial r} \left(\rho r^{4}\Omega
  U(r)\right)
  + \frac{1}{r^{2}}\frac{\partial }{\partial r}\left(\rho (D_{\rm shear}+\nu_{\rm T}) r^{4}
  \frac{\partial \Omega}{\partial r} \right) \, , 
\label{transmom}
\end{equation}
\noindent where $r$, $\rho(r)$, and $\Omega(r)$ are the radius, mean density, and mean angular velocity on an isobar, respectively. The quantity $U(r)$ corresponds to the radial dependence of the meridional circulation velocity in the radial direction and $D_{\rm shear}$ is the diffusion coefficient for AM transport by the shear instability \citep[see Sect. 2.1 of][for more details]{egg10_sl}. The transport of AM by the Tayler instability is taken into account through the viscosity $\nu_{\rm T}$ as given by Eq.~(35) of \cite{ful19}, i.e.
\begin{equation}
 \nu_{\rm T}= \alpha^{3} r^2 \Omega \left(\frac{\Omega}{N_{\rm eff}}\right)^2 \; , 
 \label{nu_T}
 \end{equation}
\noindent where $\alpha$ is a dimensionless calibration parameter of order unity. The quantity $N_{\rm eff}$ is an effective Brunt-V\"{a}is\"{a}l\"{a} frequency that accounts for the reduction of the stabilizing effect of the entropy gradient by thermal diffusion, i.e. \begin{equation}
 N^2_{\rm eff}= \frac{\eta}{K} N_T^2 + N_{\mu}^2 \; . 
 \label{Neff}
 \end{equation}The values $K$ and $\eta$ are the thermal and magnetic diffusivities, while $N_T$ and $N_{\mu}$ are the thermal and chemical composition components of the Brunt-V\"{a}is\"{a}l\"{a} frequency (with $N^2=N_T^2+N_{\mu}^2$). The minimum value of radial differential rotation needed for the magnetic AM transport to operate is given by \citep[Eq. 36 of][]{ful19}
\begin{equation}
q_{\rm min } = \alpha^{-3}\left(\frac{N_{\rm eff}}{\Omega}\right)^{5/2} \left(\frac{\eta}{r^2 \Omega}\right)^{3/4} \; ,
\label{qmin}
\end{equation}
where $q= -\frac{\partial \ln \Omega}{\partial \ln r}$. When the shear parameter $q$ is larger than the minimum threshold given by $q_{\min}$, magnetic AM transport is taken into account with the viscosity $\nu_{\rm T}$ given by Eq.~(\ref{nu_T}). The computation of this $q_{\rm min}$ condition and the viscosity $\nu_{\rm T}$ is done as described in \cite{ful19}. A very efficient AM transport is assumed in convective zones, leading to a flat rotation profile in these regions.

\section{Rotational properties of subgiant stars}
\label{results}

The present study takes place in the direct continuity of our recent work devoted to the determination of the efficiency of internal AM transport in subgiant stars (Paper I). We then computed models of the six subgiants observed by \cite{deh14} using the input parameters determined in Paper~I (the main parameters are recalled in Table~\ref{tab1} below) and following the evolution of their internal rotation profiles as described in Sect.~\ref{inputphys}. The values of the initial rotation on the zero-age MS ($V_{\rm ini}$) were determined in order to correctly reproduce the observed surface rotation rates of the subgiants. As explained in Paper I, the low values of these velocities are a direct consequence of the assumption of an inefficient magnetic braking of the stellar surface during the MS. We have shown in Paper~I that the mean efficiency of AM transport needed to reproduce the rotation rates deduced from asteroseismic measurements correctly can be precisely determined for the six subgiants, independently from their past rotational evolution (in particular regarding the modelling of AM transport and surface magnetic braking during the MS). This constitutes strong constraints that can be directly compared to the efficiency predicted by a given AM transport process in stellar radiative zones.

\begin{figure}[htb!]
\resizebox{\hsize}{!}{\includegraphics{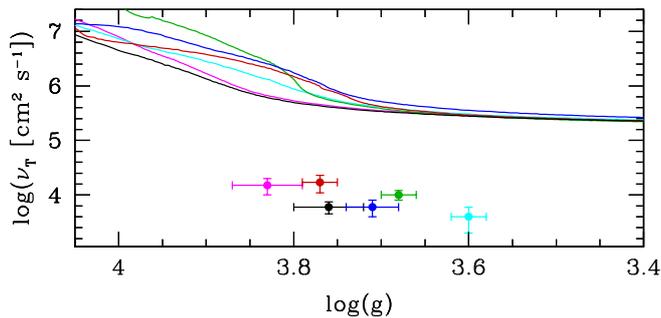}}
 \caption{Effective viscosity $\nu_{\rm T}$ associated with AM transport by the Tayler instability (Eq.~\ref{nu_T} with $\alpha=1$) evaluated at the maximum value of $N_{\mu}$ as a function of surface gravity for the models of the six subgiants studied in Paper I. The colours are the same as those used in \cite{deh14} and Paper I: magenta, red, black, blue, green, and cyan correspond to stars A, B, C, D, E, and F, respectively. 
The dots indicate the values of the mean additional viscosities determined for the six subgiants in Paper I.
}
  \label{nu_F}
\end{figure}

The viscosities associated with AM transport by the Tayler instability (Eq.~\ref{nu_T}) are shown in Fig.~\ref{nu_F} for the six subgiants studied in Paper I. The plotted values correspond to the viscosity at the border of the helium core, where the maximum value of the chemical composition part of the Brunt-V\"{a}is\"{a}l\"{a} frequency ($N_{\mu}$) is reached. The ability to transport AM in this region of strong chemical gradients directly determines the degree of radial differential rotation during the poMS evolution. As seen in Fig.~\ref{nu_F}, high values of the viscosity are obtained during the beginning of the subgiant phase. As evolution proceeds, a decrease of these viscosities is observed owing to the simultaneous increase of the Brunt-V\"{a}is\"{a}l\"{a} frequency and decrease of the rotation velocity. This trend is in qualitative agreement with the result of Paper~I about the decrease of AM transport efficiency with the evolution during the subgiant phase. However, the effective viscosities predicted by Eq.~\ref{nu_T} are much higher than the values deduced from asteroseismic measurements in Paper I, which lie between $4 \times 10^{3}$ and $1.7 \times 10^{4}$\,cm$^2$\,s$^{-1}$. This is illustrated in Fig.~\ref{nu_F}, in which the dots denote the values of the viscosities determined in Paper~I.

\begin{table}
\caption{Input parameters corresponding to the models of the six subgiants studied by \cite{deh14}.} 
\begin{center}
\label{tab:res}
\begin{tabular}{c|cccc}
\hline
\hline
Star & $M/M_{\odot}$ & $(Z/X)_{\rm ini}$ & $Y_{\rm ini}$ & $V_{\rm ini}$ [km\,s$^{-1}$] \\ \hline
A & 1.20 &  0.0550 &  0.30 & 6 \\
B & 1.27 &  0.0190 &  0.28 & 5 \\
C & 1.15 &  0.0390 &  0.29 & 5 \\
D & 1.25 &  0.0160 &  0.26 & 6 \\
E & 1.40 &  0.0500 &  0.29 & 8 \\
F & 1.10 &  0.0100 &  0.26 & 4\\
\hline
\end{tabular}
\end{center}
\label{tab1}
\end{table}

As a consequence of the high effective viscosities associated with the transport by the Tayler instability, a very low degree of radial differential rotation would be predicted during the subgiant and red giant phase. To correctly account for the asteroseismic determination of core and surface rotation rates of subgiants, we thus see that the AM transport efficiency cannot depend solely on the viscosities given by Eq.~\ref{nu_T}. We conclude that the condition on the minimum value of radial differential rotation needed for the magnetic transport mechanism to operate (Eq.~\ref{qmin}) must play the most important role in shaping the internal rotation of poMS stars as early as the subgiant phase.

\begin{figure}[htb!]
\resizebox{\hsize}{!}{\includegraphics{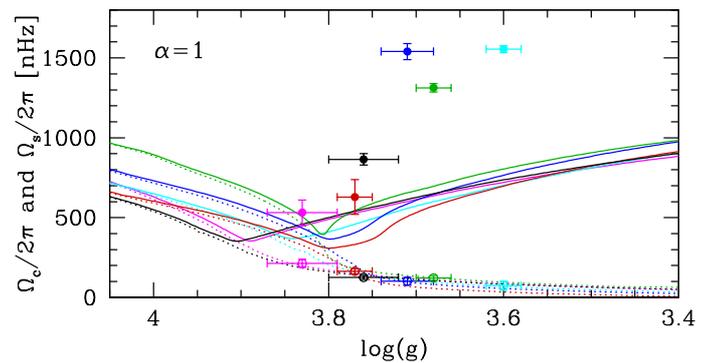}}
 \caption{Core and surface (solid and dotted lines) rotation rates as a function of gravity for models of the six subgiants. These models have been computed with a parameter $\alpha=1$. Magenta, red, black, blue, green, and cyan correspond to stars A, B, C, D, E, and F, respectively.} 
  \label{omcs_alpha1}
\end{figure}

The poMS evolution of the core and surface rotation rates of models of the six subgiants accounting for AM transport by the Tayler instability is shown as a function of the surface gravity in Fig.~\ref{omcs_alpha1}. Core rotation rates correspond to mean values in the g-mode cavity \citep{gou13}. We first computed these models with a calibration parameter $\alpha=1$. The efficient additional AM transport associated with the revised prescription for the transport by the Tayler instability leads to a low degree of radial differential rotation during the subgiant phase. Consequently, the predicted core rotation rates are lower than those deduced from asteroseismic measurements for the subgiants (see Fig.~\ref{omcs_alpha1}).

\begin{figure}[htb!]
\resizebox{\hsize}{!}{\includegraphics{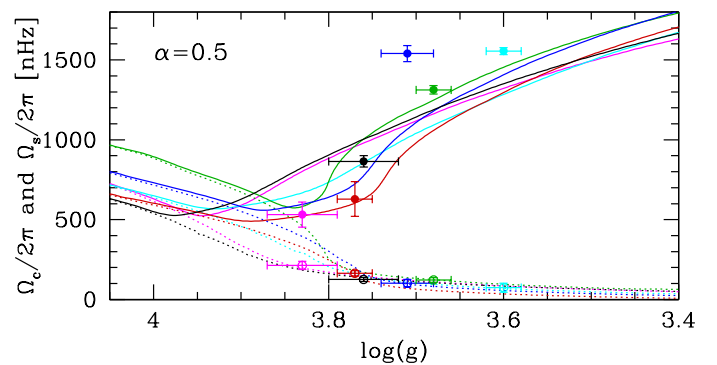}}
 \caption{Same figure as Fig.~\ref{omcs_alpha1} but for a calibration parameter $\alpha=0.5$.} 
  \label{omcs_alpha0p5}
\end{figure}

Rotating models were then computed for subgiants by varying the dimensionless calibration parameter $\alpha$. A lower value of $\alpha$ leads to a higher degree of radial differential rotation for the magnetic transport to operate (see Eqs.~\ref{nu_T} and \ref{qmin}) and thereby enabled us to better account for the core rotation rates of subgiant stars. This is shown in Fig.~\ref{omcs_alpha0p5}. With a parameter $\alpha=0.5$, core rotation rates of subgiant B (red) and E (green) can then be correctly reproduced. However, a degree of radial differential rotation that is too high is predicted for subgiants A (magenta) and C (black), while rotation rates that are too low are still obtained for subgiants D (blue) and F (cyan). Varying the values of $\alpha$, we find that at most two subgiants over six can be correctly reproduced when using the revised prescription for AM transport by the Tayler instability. This suggests that this transport process faces difficulties in reproducing the change of the transport efficiency with the mass and evolution as deduced from asteroseismic measurements of subgiant stars.

\begin{figure}[htb!]
\resizebox{\hsize}{!}{\includegraphics{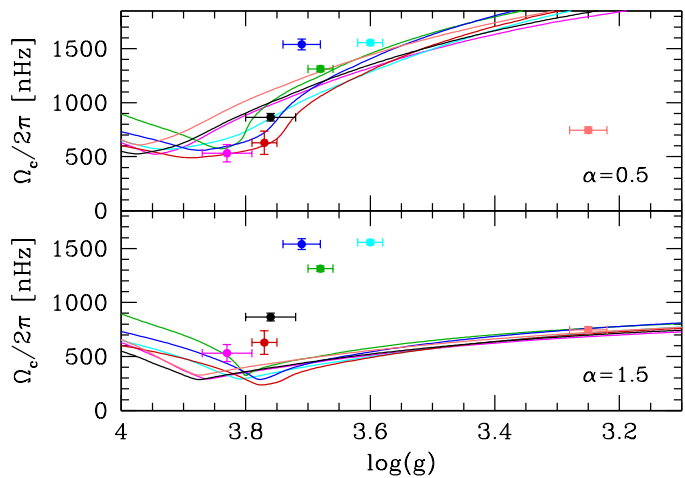}}
 \caption{Core rotation rates as a function of surface gravity for models of the six subgiants computed for different values of the calibration parameter $\alpha$. A model of the red giant KIC~4448777 is shown in orange with the corresponding value of the core rotation rate. {\it Top:} Models computed with $\alpha=0.5$, which is the value needed to reproduce correctly the core rotation rates of 2 of the 6 subgiants. {\it Bottom:} Models computed with $\alpha=1.5$, which is needed to reproduce correctly the core rotation rate of the red giant star KIC~4448777 (orange).}
  \label{omc_alphatot}
\end{figure}

We finally compared the rotational properties of models computed with the transport by the magnetic Tayler instability to asteroseismic constraints on the internal rotation of both subgiant and red giant stars. In addition to the six subgiants, we then computed a model for the more evolved red giant KIC~4448777. This red giant is particularly interesting to consider for a comparison with subgiant stars because it shares a similar value for its mass (1.1\,$M_{\odot}$) and both its core and surface rotation rates have been determined \citep{dim16}. Figure~\ref{omc_alphatot} shows the comparison between predicted and observed core rotation rates for different values of $\alpha$. For the sake of clarity, surface rotation rates are not plotted in Fig.~\ref{omc_alphatot}, but all models reproduce the constraints on the surface velocity. 

The top panel of Fig.~\ref{omc_alphatot} shows the core rotation rate for the models with $\alpha=0.5$. As discussed above, this value of $\alpha$ leads to core rotation rates in better global agreement with the asteroseismic measurements of subgiant stars. However, this low value of $\alpha$ also leads to core rotation rates that are too high during the red giant phase, as illustrated with KIC~4448777 (orange). The impact of an increase of the calibration parameter is then studied by computing models with $\alpha=1.5$. As illustrated in the bottom panel of Fig.~\ref{omc_alphatot}, this enabled us to account correctly for the core rotation rates during the red giant phase. However, a very low degree of radial differential rotation is then obtained earlier, which results in core rotation rates that are too low for subgiant stars.

\begin{figure}[htb!]
\resizebox{\hsize}{!}{\includegraphics{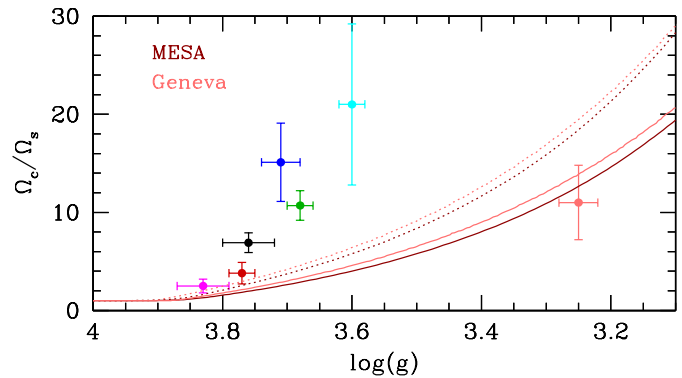}}
 \caption{Ratio of core to surface rotation rates as a function of surface gravity for models of the red giant KIC~4448777 computed with the MESA (brown lines) and the Geneva (orange lines) stellar evolution codes. The dotted and continuous lines correspond to models computed with $\alpha=1$ and $\alpha=1.5$, respectively.}
  \label{omc_jacqueline}
\end{figure}

We conclude that, whatever the value adopted for the dimensionless calibration parameter $\alpha$, the rotational properties of at most two subgiants out of six can be correctly reproduced when including the revised prescription for AM transport by the Tayler instability proposed by \citet[][]{ful19}. Moreover, a value of the parameter $\alpha$ calibrated to reproduce the core rotation rate of the red giant KIC~4448777 (orange dot in Fig.~\ref{omc_alphatot}) leads to a very low degree of radial differential rotation in subgiants, which contrads with the asteroseismic constraints on the internal rotation available for these stars.

These results have been obtained from asteroseismic models computed with the Geneva stellar evolution code. We investigated the robustness of these results compared to the use of a different evolution code (that can differ in the input physics and the numerical methods used) using the MESA code \citep[][]{pax11,pax13,pax15,pax18,pax19}. Models were then computed with the MESA code (revision 11701) with the same implementation of AM transport as described in \citet[][]{ful19}. Models of the red giant KIC~4448777 computed with both codes with similar input parameters are shown in Fig.~\ref{omc_jacqueline}.  We find that similar rotational properties are obtained with the MESA and Geneva codes. Whichever the evolution code is used, models that correctly reproduce the core rotation rates of red giants always predict an AM transport that is too efficient before the red giant phase to account correctly for the asteroseismic constraints available for subgiants. This issue is directly related to the strong decrease in the AM transport efficiency predicted by the revised magnetic process during the red giant phase, which is due to both a decrease in the viscosity $\nu_{\rm T}$ (see Fig.~\ref{nu_F}) and an increase in the $q_{\rm min}$ parameter. To obtain an efficient AM transport that is able to reproduce the core rotation rates of red giants, this mechanism then leads to an even more efficient transport before the red giant phase; this is found to be in contradiction with the radial differential rotation observed in subgiants (see Fig.~\ref{omc_jacqueline}).

\section{Conclusions}
\label{conclusion}

Based on our previous characterisation of internal AM transport during the subgiant phase (Paper I), we confront rotating models that account for magnetic AM transport as recently proposed by \citet[][]{ful19} to these asteroseismic constraints. 

We first show that the condition on the minimum radial differential rotation needed for the transport by the Tayler instability to operate plays a major part in shaping the rotation profile of the star as early as the subgiant phase. This result is similar to that obtained for more evolved red giant models by \citet[][]{ful19}. The core rotation rates predicted for subgiants by this revised prescription are also found to be much lower and thus in better global agreement with asteroseismic measurements of subgiants than those predicted by models with purely hydrodynamic processes or with the original Tayler-Spruit dynamo.

A detailed comparison between models accounting for the magnetic AM transport proposed by \citet[][]{ful19} and the asteroseismic constraints available for the six subgiants studied in Paper~I shows that the rotational properties of at most two of the six stars can be correctly reproduced. This result is obtained independently from the value adopted for the dimensionless calibration parameter $\alpha$ introduced in the expression for the AM transport. This indicates that the functional dependence of the revised condition expressing the minimum shear required for the magnetic transport to operate is not fully compatible with the asteroseismic constraints on the internal rotation of subgiant stars.

We also find that the revised magnetic transport process faces difficulties in simultaneously reproducing the core rotation rates observed in red giant and subgiant stars. The low values of the calibration parameter $\alpha$ needed to account correctly for the rotational properties of two of the six subgiants result indeed in core rotation rates that are too high during the red giant phase. Conversely, the higher values of $\alpha$ needed to reproduce the core rotation rates of red giants lead to a very low degree of radial differential rotation before the red giant phase, which contradicts with the asteroseismic measurements available for subgiant stars.

These results underline the difficulty of finding a physical process that is able to reproduce correctly the asteroseismic measurements available for poMS stars. It is indeed particularly difficult to account simultaneously for the decrease in the AM transport efficiency observed during the subgiant phase \citep[][]{egg19} and then the increase of this transport efficiency needed when the star ascends the red giant branch \citep[][]{egg17}. This also illustrates the fundamental role played by asteroseismic measurements of evolved stars to test and try to improve the modelling of the different physical processes for AM transport in stellar radiative zones.

\begin{acknowledgements}
We would like to thank the referee, Dr. Jim Fuller, for the useful comments that helped us improve the quality of the paper.
This work has been supported by the Swiss National Science Foundation (project Interacting Stars, number 200020-172505). JWdH acknowledges funding by ERC-2016-CO Grant 724560 and “ChETEC” COST Action (CA16117). 
\end{acknowledgements}


\bibliographystyle{aa} 
\bibliography{biblio} 

\end{document}